\documentclass[journal=jacsat,manuscript=article]{achemso}

\usepackage{chemformula} 
\usepackage[T1]{fontenc} 
\usepackage{color,soul}

\usepackage{multirow}

\author{Edoardo Trabaldo}
\affiliation[Charmers University of Technology]{Quantum Device Physics Laboratory, Department of Microtechnology and Nanoscience, Chalmers University of Technology, SE-41296 G\"{o}teborg, Sweden}
\author{Christoph Pfeiffer}%
\affiliation[Charmers University of Technology]{Quantum Device Physics Laboratory, Department of Microtechnology and Nanoscience, Chalmers University of Technology, SE-41296 G\"{o}teborg, Sweden}
\author{Eric Andersson}%
\affiliation[Charmers University of Technology]{Quantum Device Physics Laboratory, Department of Microtechnology and Nanoscience, Chalmers University of Technology, SE-41296 G\"{o}teborg, Sweden}
\author{Riccardo Arpaia}%
\affiliation[Charmers University of Technology]{Quantum Device Physics Laboratory, Department of Microtechnology and Nanoscience, Chalmers University of Technology, SE-41296 G\"{o}teborg, Sweden}
\alsoaffiliation[Second University]
{Dipartimento di Fisica, Politecnico di Milano, Piazza Leonardo da Vinci 32, I-20133 Milano, Italy}
\author{Alexei Kalaboukhov}%
\affiliation[Charmers University of Technology]{Quantum Device Physics Laboratory, Department of Microtechnology and Nanoscience, Chalmers University of Technology, SE-41296 G\"{o}teborg, Sweden}
\author{Dag Winkler}%
\affiliation[Charmers University of Technology]{Quantum Device Physics Laboratory, Department of Microtechnology and Nanoscience, Chalmers University of Technology, SE-41296 G\"{o}teborg, Sweden}
\author{Floriana Lombardi}%
\affiliation[Charmers University of Technology]{Quantum Device Physics Laboratory, Department of Microtechnology and Nanoscience, Chalmers University of Technology, SE-41296 G\"{o}teborg, Sweden}
\author{Thilo Bauch}%
\affiliation[Charmers University of Technology]{Quantum Device Physics Laboratory, Department of Microtechnology and Nanoscience, Chalmers University of Technology, SE-41296 G\"{o}teborg, Sweden}
\email{thilo.bauch@chalmers.se}

\title[]
  {Grooved Dayem nanobridges as building blocks of high-performance YBa$_2$Cu$_3$O$_{7-\delta}$ SQUID magnetometers}

\begin{document}

\begin{abstract}
  We present noise measurements performed on a YBa$_2$Cu$_3$O$_{7-\delta}$ { nanoscale} weak-link-based magnetometer consisting of a Superconducting QUantum Interference Device (SQUID) galvanically coupled to a $3.5 \times 3.5~$mm$^2$ pick-up loop, reaching white flux noise levels and magnetic noise levels as low as $6~\mu\Phi_0 / \sqrt{\mathrm{Hz}}$ and $100$~fT/$\sqrt{\mathrm{Hz}}$ at $T=77$~K, respectively. The low noise is achieved by introducing Grooved Dayem Bridges (GDBs), a new concept of weak-link. A fabrication technique has been developed for the realization of nanoscale grooved  bridges, which substitutes standard Dayem bridge weak links. The introduction of these novel key blocks reduces the parasitic inductance of the weak links and increases the differential resistance of the SQUIDs. This greatly improves the device performance, thus resulting in a reduction of the white noise.
\end{abstract}

Superconducting QUantum Interference Devices (SQUIDs) are among the most sensitive magnetometers available today, making them one of the most prominent devices for various applications of superconducting materials. Since their introduction, SQUIDs have been used in several technological applications, e.g. geophysical surveys \cite{clarke1983geophysical}, medical diagnostic (MCG and MEG) \cite{koch2001squid,oisjoen2010new,xie2017improved} and scanning SQUID microscopy \cite{vasyukov2013scanning}. 

While the basic operation of SQUIDs is well established \cite{clarke_handbook_2006}, great effort is still invested in the improvement of their performance. In this respect, the recent technological advances in nano-fabrication enabled the realization of nanoSQUIDs with white flux noise levels well below $1$~$\mu\Phi_0/\sqrt{\mathrm{Hz}}$ ~\cite{granata2013three, arpaia2014ultra, wolbing2014optimizing, schwarz2015low, arzeo2016toward, arpaia2016improved, chen2016high, russo2016nanosquids}, opening the way to single spin detection, a milestone of experimental physics \cite{martinez2016nanosquids, granata2016nano}.

SQUIDs made of High critical Temperature Superconductors (HTS) have a much wider temperature range of operation (from mK to above 77~K) compared to their Low critical Temperature Superconductor (LTS) counterparts, greatly simplifying their practical applications. HTS SQUID magnetometers are promising candidates for future on-scalp magnetoencephalography  systems \cite{schneiderman2014information,xie2017benchmarking,riaz2017evaluation,oisjoen2012high}. Tremendous efforts have been devoted to achieve high quality HTS Josephson Junctions (JJs), the key ingredient of a SQUID, during the last few decades. This has proven to be challenging for cuprate HTS materials, due to the chemical instability, the small superconducting coherence length ($\sim 2$~nm in the ab-planes) {and the ceramic and granular nature} of these materials. Nevertheless, different JJ fabrication techniques have been successfully developed for HTS SQUIDs so far. For example, high sensitivity HTS SQUIDs have been realized using grain boundary based JJs, by epitaxial growth of HTS films on bicrystal or step edge substrates  \cite{lee1995key,faley2013high,oisjoen2012high,chukharkin2013improvement, schwarz2015low,chesca2015flux,mitchell2010ybco}. 

HTS nanoSQUIDs, realized with Dayem bridges, have also shown low magnetic flux noise properties, in combination with a simplified fabrication procedure \cite{arpaia2014ultra,arzeo2016toward, arpaia2016improved}. However, the rather large parasitic inductance of Dayem bridges limits their implementation in SQUID magnetometers at $77~$K \cite{xie2017improved}. {  Another approach to fabricate HTS weak links is by high energy ion irradiation (30-200~keV) of predefined wide YBCO bridges through a mask \cite{tinchev2007mechanism} or by direct irradiation using a He focused ion beam (FIB) \cite{cybart2015nano}. This resulted in high quality low noise SQUID devices at 4~K and 50~K\cite{cho2015yba2cu3o7,cho2018direct}. RF-SQUIDs working at 77~K and above have been realized showing a white flux noise level of $100-200$~$\mu\Phi_0  / \sqrt{\mathrm{Hz}}$ \cite{tinchev1993high}, which is more than one order of magnitude higher than values obtained in state of the art dc-SQUIDs \cite{oisjoen2012high,chukharkin2013improvement,faley2014graphoepitaxial,faley2013high,mitchell2010ybco}. DC-SQUID devices operational at 77~K have been realized by oxygen ion irradiation, however no frequency dependent noise data have been reported \cite{bergeal2006high}.}

In this work we present a novel fabrication process of a HTS weak link: the nanoscale Grooved Dayem Bridge (GDB), which exhibits Josephson Junction-like behavior. Here, the layout of the bridge and the weak link inside the bridge are realized during one single lithography process on a YBa$_2$Cu$_3$O$_{7-\delta}$ (YBCO) film grown on a single crystal substrate. Moreover, such weak links can be defined anywhere on the chip and freely oriented within the film plane. This approach has clear advantages compared to bicrystal HTS junctions, where the junctions are located at the grain boundary line, and to step edge junctions, which involve more than one lithography step \cite{foley1999fabrication} and several epitaxial thin film depositions \cite{faley2014graphoepitaxial}.

We have used YBCO GDBs as novel, alternative building { nanoscale} blocks in HTS SQUID magnetometers, which have been characterized via transport and noise measurements at T$=77$~K. In particular, these devices exhibit large voltage modulations ($\Delta V = 30-50~\mu$V) and low values of white magnetic flux noise, $6~\mu\Phi_0/\sqrt{\mathrm{Hz}}$ { (compared to the lowest reported flux noise $4.5~\mu\Phi_0/\sqrt{\mathrm{Hz}}$ at $T=77$~K \cite{mitchell2010ybco})}, and corresponding magnetic field noise, $100~$fT$/\sqrt{\mathrm{Hz}}$, at $T=77$~K. Therefore, GDB based SQUIDs combine the nanofabrication advantages and the device reproducibility, which are typical of Dayem bridges, with the performances, e.g. the  magnetic sensitivity, of state-of-the-art SQUIDs based on grain boundary JJs.

\begin{figure}[]
\includegraphics[width=0.485\textwidth]{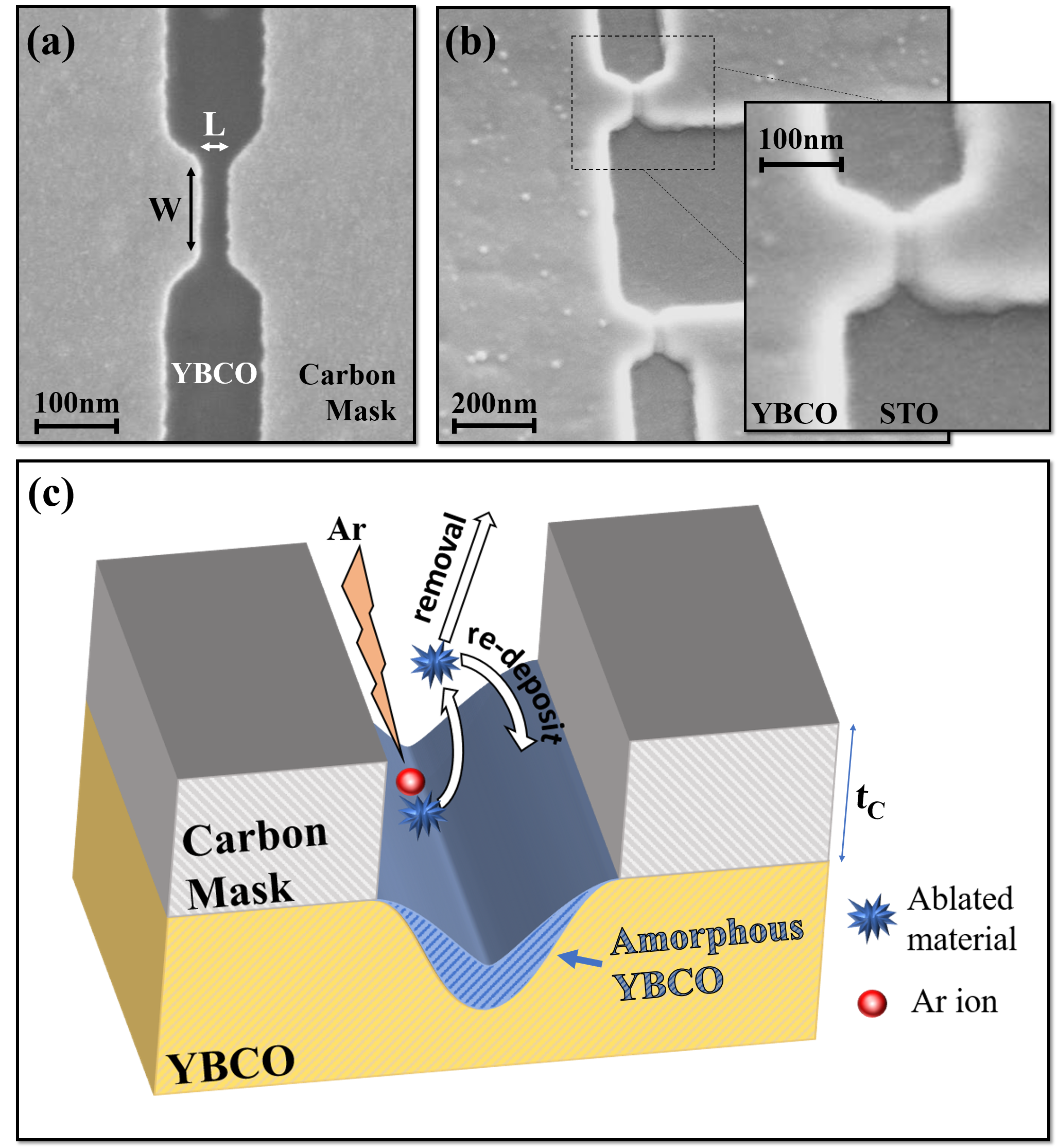}
\caption{(a) Scanning electron microscope (SEM) image of the carbon mask on top of the YBCO film. The width and length of the gap are indicated by  $W$ and $L$, respectively. (b) SEM image of two GDBs after ion milling and removal of the carbon mask. The inset is the zoom-in of a single GDB. (c) Schematic of the Ar ion milling process inside the gap. The ion milled YBCO is partially redeposited as amorphous material inside the GDB gap.}
\label{Fig:Mask}
\end{figure}

A $50$~nm thick film of YBCO is deposited by Pulsed Laser Deposition (PLD) onto a (001) SrTiO$_3$ substrate with lateral dimensions of $5\times5$~mm$^2$. On top of the YBCO film, a mask layer of amorphous hard carbon with thickness $t_C=110$~nm is deposited by PLD and defined by e-beam lithography (EBL). The device is finally patterned by gentle { low energy (300~eV)} Ar ion milling, as described in previous works \cite{nawaz2013microwave, nawaz2013approaching}. { The detailed ion milling parameters are summarized in the supporting information, table S1.} This fabrication procedure has been shown to result in YBCO nanowires with pristine bulk-like properties \cite{nawaz2013microwave,nawaz2013approaching,trabaldo2017noise, arpaia2018probing}. Here, in order to fabricate the Grooved Dayem Bridges, we start from this fabrication and take advantage of a reduced etching rate of the ion milling in specifically designed areas of the sample. To achieve this, the standard mask design for the Dayem bridge is modified, opening a gap along the full width, as shown in Figure \ref{Fig:Mask}(a).

The width $W$ and length $L$ of the gap define the geometrical dimensions of the final GDB and can be varied to achieve different values of critical current $I_{\mathrm{C}}$. For aspect ratios of the gap in the carbon mask $t_C/L>2$, the etching rate of YBCO during the Ar ion milling inside the gap is strongly reduced compared to the rest of the sample. This is the result of partial re-deposition of the YBCO ablated by the Ar ions, which cannot be removed from the gap and forms an amorphous layer of YBCO \cite{manos1989plasma} (see Fig.\ref{Fig:Mask}(c)). Since the etching rate inside the gap is reduced, the ion milling process required to remove $50$~nm of YBCO far away from the carbon mask edges results only in a slight thinning of the YBCO inside the gap, leaving a grooved bridge with thickness less than $50$~nm. The final result obtained for a SQUID after the final ion milling is shown in Figure \ref{Fig:Mask}(b).

\begin{figure}[]
\includegraphics[width=0.48\textwidth]{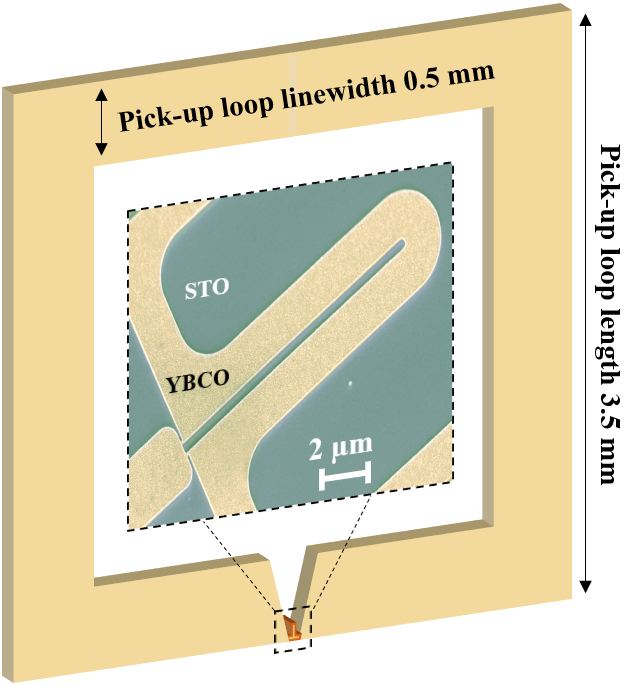}
\caption{Schematic of the pick-up loop, with the dimensions used for the device SQ1. The inset is an SEM image, showing the details of the SQUID loop of the device SQ2, with hole dimensions $12\times0.5$~$\mu$m$^2$. The line width of the hairpin loop is 2~$\mu$m.}
\label{Fig:Pickuploop}
\end{figure}

Up to 16 GDB-SQUIDs can be fabricated on the same pick-up loop. The widths of the GDBs are $150 - 200~$nm and the gap length in the bridge mask $50~$nm. All the SQUIDs are galvanically connected to the same in plane square shaped pick-up loop of lateral dimension $3.5 \times 3.5~$mm$^2$ in order to increase the effective magnetic area $A_{\mathrm{eff}}=\Phi/B_a$, with $\Phi$ the magnetic flux through the SQUID loop and $B_a$ the externally applied magnetic field \cite{arzeo2016toward,xie2017improved}. A schematic of the pick-up loop, with only a single SQUID coupled to it, is shown in Figure \ref{Fig:Pickuploop}.

When an external magnetic field $B_a$ is applied, a screening current $I_{\mathrm{S}}\propto B_a$ circulates in the pick-up loop. $I_{\mathrm{S}}$ generates a phase difference $\Delta \phi$ between the two weak links, which is proportional to $I_{\mathrm{S}}\cdot  L_{\mathrm{c}} \cdot 2\pi / \Phi_0$, where $\Phi_0$ is the superconducting flux quantum. Here $L_{\mathrm{c}}$ is the coupling inductance shared between the pick-up loop and the SQUID loop, i.e. the hairpin loop shown in the inset of Fig. \ref{Fig:Pickuploop}. For the resulting effective area of the SQUID $A_{eff} = \Phi_0 /2\pi \cdot \Delta\phi /B_a$ one can write \cite{arzeo2016toward}

\begin{equation}
A_{\mathrm{eff}}=A_{\mathrm{nS}}+A_{\mathrm{eff}}^{\mathrm{pl}}\frac{L_{\mathrm{c}}}{L_{\mathrm{loop}}} ,
\end{equation}

where $A_{\mathrm{nS}}$ is the effective area of the SQUID loop, $A_{\mathrm{eff}}^{\mathrm{pl}}$ is the effective area of the pick-up loop and $L_{\mathrm{loop}}$ is the inductance of the pick-up loop. A more detailed analysis of the effect of a pick-up loop on the SQUID performance can be found in Refs. \cite{arzeo2016toward,xie2017improved}. $L_{\mathrm{c}}$ has been measured in a separate experiment by direct current injection and it is directly proportional to the length of the hairpin slit $l_{\mathrm{slit}}$. It can be approximated from the measured data as $L_{\mathrm{c}}\simeq l_{\mathrm{slit}}\cdot 8$~pH/$\mu$m at $T=77$~K and is mainly dominated by kinetic inductance $L_k\simeq 2l_{slit}\mu_0\lambda_L^2/w_{hl} t$, with $\mu_0$ the vacuum permeability, $w_{hl}=2~\mu$m the line width of the hair pin loop, $t$ the thickness of the YBCO film, and $\lambda_L\simeq 565~$nm the in plane London penetration depth at 77~K \cite{arzeo2016toward}.

\begin{table*}[]
\begin{tabular}{c c c c c c c c c c c}
\hline
\hline
\multirow{2}{*}{{Name}} & \multirow{2}{*}{{Coupling}}  & {$l_{\mathrm{slit}}$} & {$W$} & {$L$} & {$I_\mathrm{C}$} & {$\Delta V$} & $L_\mathrm{C}$ & $A_{\mathrm{eff}}$ & $I_{\mathrm{C}} \delta R$ & $I_{\mathrm{C}} R_\mathrm{N}$ \\ 
 & & [$\mu$m] & [nm] & [nm] & [$\mu$A] & [$\mu$V] & [pH] & [mm$^2$] & [$\mu$V] & [$\mu$V]\\ 
 \hline
 \hline
 \multirow{2}{*}{SQ1} &\multirow{2}{*}{ Pick-up loop} & \multirow{2}{*}{20} & \multirow{2}{*}{200} & \multirow{2}{*}{50} & \multirow{2}{*}{15} & \multirow{2}{*}{30} & \multirow{2}{*}{160} & \multirow{2}{*}{0.122} & \multirow{2}{*}{187} & \multirow{2}{*}{346} \\
& & & & & & & & & &\\
 \hline
 \multirow{2}{*}{SQ2} &  Direct current & \multirow{2}{*}{12} & \multirow{2}{*}{150} & \multirow{2}{*}{50} & \multirow{2}{*}{7} & \multirow{2}{*}{50} & \multirow{2}{*}{96} & \multirow{2}{*}{-} & \multirow{2}{*}{168} & \multirow{2}{*}{364}   \\
& injection & & & & & & & & &\\
\hline
 \end{tabular}
\caption{Summary of the SQUIDs geometrical and transport parameters at $T=77~$K.}
\label{tab:my_label}
\end{table*}

The GDB-SQUIDs have been characterized via electrical transport and noise measurements, performed in a magnetically shielded room { (shielding factor $10^2-10^5$ in the frequency range $0.1~$Hz$-1~$kHz)} at liquid nitrogen temperature, $T\simeq 77$~K. A summary of the geometric dimensions and electric transport properties of two SQUIDs patterned on two different chips are reported in Table~1. SQ2 didn't have a pick-up loop but was measured with direct current injection. Here, a modulation current $I_\mathrm{m}$ is applied directly to the SQUID loop, which acts as an effective magnetic flux $I_\mathrm{m}\times L_\mathrm{c}$ , therefore inducing modulations of critical current $I_{\mathrm{C}}$ \cite{johansson2009properties}.

\begin{figure}[]
\includegraphics[width=1\textwidth]{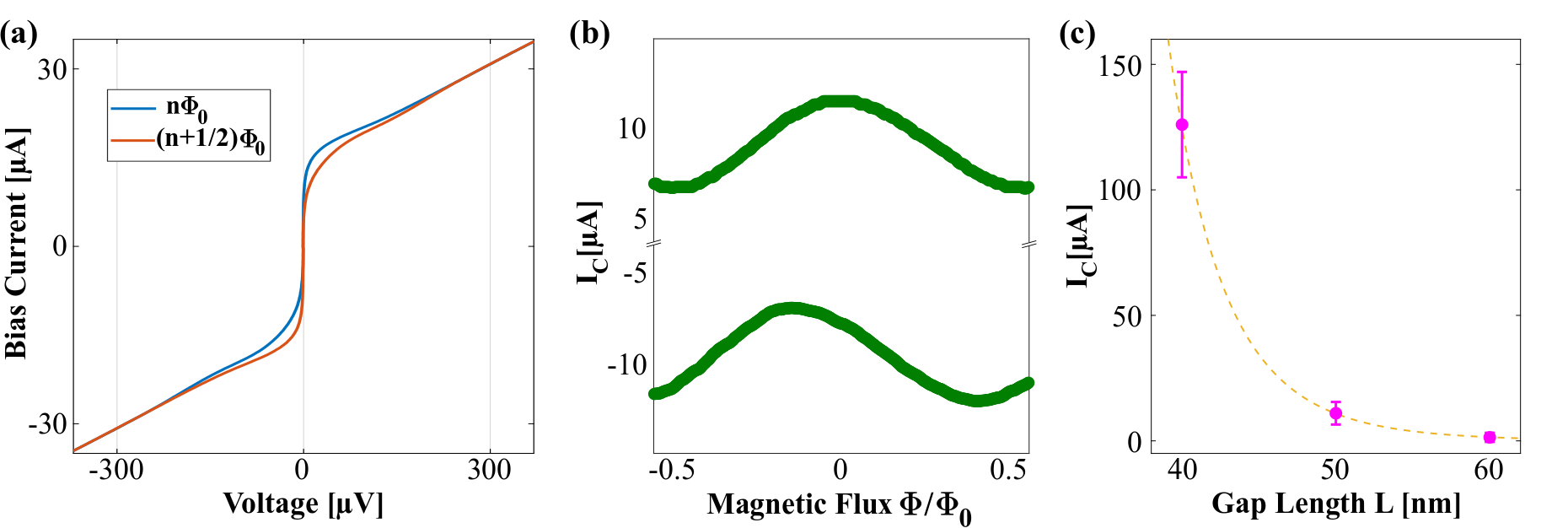}
\caption{(a) Current Voltage Characteristics (IVCs) of SQUID SQ1 at 77~K. The red and blue curves correspond respectively to the maximum and to the minimum value of the positive critical current $I_\mathrm{C}$ within one modulation period. The asymmetry in the modulation between positive and negative bias currents is a consequence of the used asymmetric biasing scheme. (b) Modulation of $I_\mathrm{C}$ as a function of the applied magnetic flux. (c) Average $I_\mathrm{C}$ measured on several GDBs for various gap lengths $L$ and fixed bridge width $W=200~$nm. The error bars are obtained as the standard deviations of the $I_\mathrm{C}$ measured on an ensemble of devices. Over 9 different fabricated chips the total amount of measured devices are $28$, $56$ and $16$ for $L=40$, $50$ and $60~$nm respectively. The dashed line is a guide for the eye.}
\label{Fig:IVC}
\end{figure}

The two current voltage characteristics (IVCs) shown in Figure \ref{Fig:IVC}(a) for SQ1 correspond to the maximum, $I_{\mathrm{C}}^{\mathrm{max}}$, and the minimum, $I_{\mathrm{C}}^{\mathrm{min}}$, measured values of the positive critical current within one modulation period. Figure \ref{Fig:IVC}(b) shows the modulation of the critical current, $I_{\mathrm{C}}(\Phi)$, where $\Phi$ is the applied magnetic flux. Similarly to what has been shown in our previous works on Dayem bridge SQUIDs \cite{arpaia2014ultra,arzeo2016toward, arpaia2016improved}, our GDB based SQUIDs exhibit modulations of the critical current $I_{\mathrm{C}}$ as a function of the externally applied magnetic flux. The IVCs and $I_{\mathrm{C}}$ modulation for SQ2 are reported in the supporting information showing similar behavior as SQ1.

The shape of the IVC resembles that of a resistively shunted junction (RSJ) similar to junctions defined with a focused helium ion beam \cite{cybart2015nano}, indicative of a { superconductor-normal conductor-superconductor (SNS) type of junction at $77~$K. The RSJ-like behavior persists down to $T=4.2~$K (see supporting information), a feature not observed in ,e.g., oxygen irradiated bridges \cite{bergeal2006high}.} Here it is important to point out that the characteristic voltage of GDBs $I_{\mathrm{C}}R_{\mathrm{N}}\simeq 350~\mu$V, with $R_{\mathrm{N}}$ the differential resistance taken in the voltage range $V = 300 - 500~\mu$V, is at least a factor of 10 larger than { He} ion irradiated junctions reported in literature at $77~$K \cite{cybart2015nano}. { The highest $I_{\mathrm{C}}R_{\mathrm{N}}$ products at liquid nitrogen temperature have been achieved with grain boundary junctions \cite{faley2017high,poppe2001properties}, with  characteristic voltages in the range $800-1200~\mu$V. Bicrystal and step-edge JJs are generally believed to be closer to tunnel-like superconductor-insulator-superconductor (SIS) JJs and they are expected to have higher $I_{\mathrm{C}}R_{\mathrm{N}}$ products compared to SNS-type JJs. While the characteristic voltages of GDBs presented here are lower than for grain boundary junctions, GDB-based SQUID magnetometers are  in the same performance range as those realized by state-of-the-art junction technologies as will be shown below. A more in depth analysis on the properties of GDBs is beyond the scope of this letter and will be reported elsewhere.}

The reproducibility of the fabrication process introduced above is supported by the small spread of the measured $I_{\mathrm{C}}$ values at $T=77~$K of Grooved Dayem Bridges patterned on different chips, see Fig. \ref{Fig:IVC}(c). Here the gap length $L$ is varied between $40$ and $60$~nm and the bridge width is kept fixed $W=200~$nm. For $L=40$~nm, the average $I_{\mathrm{C}}$ is $130\pm20~\mu$A and for $L=50$~nm we measured an average $I_{\mathrm{C}}$ of $10\pm4~\mu$A. When $L$ is increased to $60$~nm, $I_{\mathrm{C}}$ is suppressed to a few $\mu$A or in most cases no supercurrent could be detected. After several cooling cycles and more than $3$ months storage at room temperature, the values of $I_{\mathrm{C}}$ do not significant deviate from the original results.

\begin{figure}[]
\includegraphics[width=0.5\textwidth]{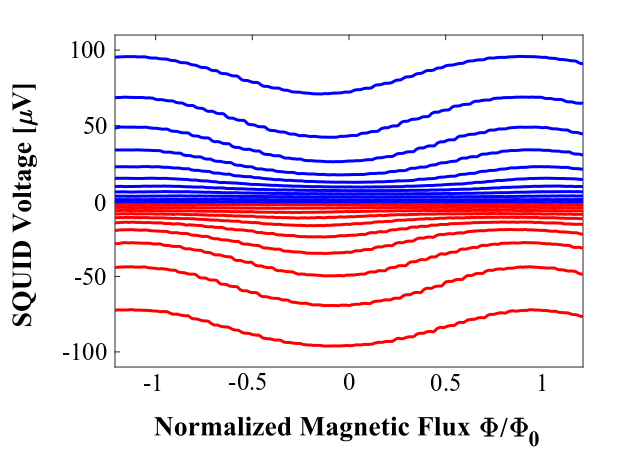}
\caption{Voltage modulations of { SQUID} SQ1 as a function of the magnetic flux for increasing values of the bias current.}
\label{Fig:V_Phi}
\end{figure}
 
In the following we discuss the advantages of using GDB-based SQUIDs compared to those implementing Dayem bridges: 
1) the critical current density of $2.5\cdot 10^5$~A/cm$^2$ is 40 times smaller than those of Dayem bridges. This allows implementing wider GDB junctions in { SQUID} magnetometers resulting in a smaller parasitic inductance, which scales with $\lambda_L^2/Wt$. Hence a larger coupling inductance (hair pin loop) can be used for increasing the effective area \cite{xie2017improved} while keeping the screening parameter near the optimal value $\beta_L = I_{\mathrm{C}}\cdot L_{\mathrm{sq}}/\Phi_0\simeq 1$, where $L_{\mathrm{sq}}$ is the total SQUID loop inductance; 2) the resistivity of GDBs is approximately $0.15~\Omega\mu$m$^2$, which is roughly 50 times larger than those of Dayem bridges. The larger resistivity of GDB results in an increased voltage modulation depth, reducing the contribution of the readout electronics input noise to the total flux noise, which will be discussed in the following.

The SQUID noise has been measured using a commercial Magnicon SEL-1 dc SQUID electronics \cite{magnicon}. The system allows to perform measurements in Flux Locked Loop (FLL) mode and use current bias reversal at $40$~kHz in order to reduce the low-frequency critical current noise \cite{drung2003high}. 

The measured voltage modulations of SQ1 as a function of magnetic flux are shown in Figure \ref{Fig:V_Phi} for various bias current values. Each curve corresponds to an increment of the bias current equal to $1.8$~$\mu$A. We obtain a voltage modulation depth $\Delta V\simeq30$~$\mu$V for values of the bias current slightly above $I_{\mathrm{C}}$. For the transfer function $V_\Phi$ defined as max$(\delta V/ \delta \Phi)\simeq \pi\Delta V/\Phi_0$, corresponding to the maximum voltage response of the SQUID per unit of flux, we obtain $V_\Phi =115$~$\mu$V/$\Phi_0$. Indeed, a large voltage modulation depth $\Delta V \simeq \delta R \Delta I_\mathrm{C}$, with $\Delta I_{\mathrm{C}}$ the critical current modulation depth and $\delta R$ the differential resistance of the SQUID at the working point (see supporting information) is required to achieve low noise SQUID devices. By doing so, the contribution of the amplifier input voltage noise $S_{V,a}^{1/2}$ to the white flux noise $S_{\Phi ,a}^{1/2}\simeq \Phi_0 S_{V,a}^{1/2}/\pi\Delta V$ of the { SQUID} is minimized, hence improving the device performance.

\begin{figure}[]
\includegraphics[width=0.48\textwidth]{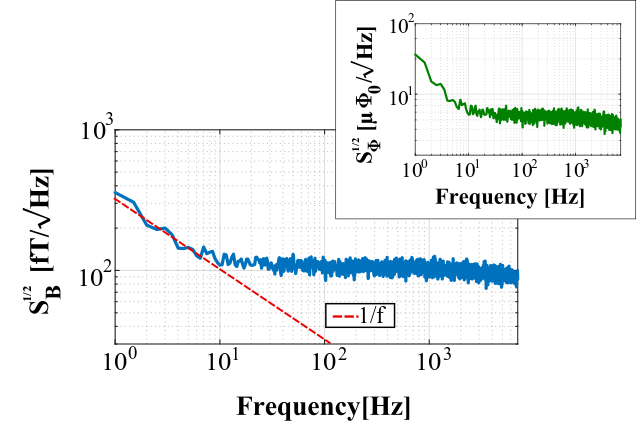}
\caption{Magnetic field noise $S_{\mathrm{B}}$ measured on SQ1 with bias reversal scheme. The red line indicates the 1/f noise. Inset: magnetic flux noise $S_\Phi$.}
\label{Fig:Noise_PL}
\end{figure}

The effective area $A_{\mathrm{eff}}$ was determined via responsivity measurements, i.e. measuring the SQUID response in FLL mode to a known magnetic field. The value of the effective area for SQ1, which is galvanically coupled to a pick-up loop with dimensions $3.5\times 3.5 $~mm$^2$, is $A_\mathrm{eff}=0.122$~mm$^2$. This value is consistent with previously reported measurements performed on similar devices implementing YBCO Dayem bridges \cite{xie2017improved}. The voltage noise $S_{\mathrm{V}}^{1/2}$ for { SQUID} SQ1 was measured in FLL mode using current bias reversal. The resulting magnetic flux noise $S_\Phi^{1/2}=S_{\mathrm{V}}^{1/2}/V_\Phi$ and magnetic field noise $S_B^{1/2}=S_\Phi^{1/2}/A_{\mathrm{eff}}$ are shown in Figure \ref{Fig:Noise_PL}. 

We observe a white flux noise level $S_\Phi^{1/2} \simeq 6$~$\mu\Phi_0/\sqrt{\mathrm{Hz}}$ above the $1/f$ knee frequency of $f\simeq 10~$Hz (see inset of Figure \ref{Fig:Noise_PL}). This noise value is close to the expected readout electronics input noise, $4$~$\mu\Phi_0/\sqrt{\mathrm{Hz}}$. The corresponding white magnetic noise is $S_B^{1/2} \simeq 100$~fT$/\sqrt{\mathrm{Hz}}$ (see Figure \ref{Fig:Noise_PL}). { This is a high level of sensitivity for a device with a much smaller pick-up loop ($3.5 \times 3.5~$mm$^2$) than standard magnetometers ($8 \times 8~$mm$^2$ and above \cite{xie2017improved,faley2017high}).} Comparing these values to Dayem bridge based magnetometers \cite{xie2017improved} GDB-based devices have 10 times lower flux and magnetic field noise. This is even more remarkable considering the much smaller pick-up loop size used in this work compared to the one used in Ref. \cite{xie2017improved}.  

The achieved white magnetic flux noise values are comparable to state of the art { single layer} grain boundary junction based YBCO SQUIDs, $S_\Phi^{1/2}= 4.5 - 10~ \mu\Phi_0/\sqrt{\mbox{Hz}}$ \cite{oisjoen2012high,chukharkin2013improvement,faley2014graphoepitaxial,faley2013high,mitchell2010ybco}.  Single layer grain boundary based SQUIDs galvanically coupled  to a pick-up loop of $\sim 8\times8$~mm$^2$ have reached white magnetic field noise values down to $S_B^{1/2}= 30-50$~fT$/\sqrt{\mathrm{Hz}}$ \cite{oisjoen2012high,chukharkin2013improvement,faley2014graphoepitaxial}. These values are a factor $2-3$ lower than the one shown in Figure \ref{Fig:Noise_PL}. However, considering the small size of the used pick-up loop, $3.5 \times 3.5$~mm$^2$, implemented in this work, grooved Dayem bridge based magnetometers should reach similar or even better magnetic field sensitivities for pick-up loops of lateral dimension $8-10$~mm. { The effective area of a magnetometer, and hence the magnetic field noise, can be further improved by implementing a superconducting flux transformer in a flip-chip setup \cite{faley2017high}. Values below $10$~fT$/\sqrt{\mathrm{Hz}}$ have been achieved in grain boundary JJ based devices \cite{faley2017high}. Implementing a flux transformer in our GDB based SQUID devices would indeed further improve the magnetic field sensitivity.}
 
In conclusion, we have developed a reproducible nanopatterning procedure for the realization of YBCO Grooved Dayem Bridges. Here the layout of the bridge and the weak link across it are realized during one single lithography process on a YBCO film grown on a single crystal substrate. Such weak links have been implemented in { SQUIDs} galvanically coupled to a square shaped in-plane pick-up loop with lateral dimension of $3.5$~mm. The smaller critical current densities and larger differential resistances of the GDBs compared to bare Dayem bridges allows to implement wider bridges in SQUID application and therefore reducing the parasitic inductance. We obtained a voltage modulation depth $\Delta V\simeq 30$~$\mu$V for a { SQUID} with hair pin slit length $l_{\mathrm{slit}}=20~\mu$m and an effective area $A_{\mathrm{eff}}=0.122~$mm$^2$.

The achieved magnetic field noise of $100~$fT$/\sqrt{\mathrm{Hz}}$ on such a small device ($3.5 \times 3.5~$mm$^2$) paves the ground for the realization of a single layer YBCO magnetometer with magnetic field noise below $25~$fT$/\sqrt{\mathrm{Hz}}$. This could be achieved on a $10\times 10~$mm$^2$ substrate and using a slightly longer hair pin SQUID loop. This work proves the feasibility of Grooved Dayem Bridges for the fabrication of high quality weak links and for SQUID applications. { The development of low noise HTS SQUIDs is crucial not only for technological applications, such as medical diagnostic (MCG and MEG) \cite{koch2001squid,oisjoen2010new,xie2017improved} and geophysical surveys \cite{clarke1983geophysical}, but also in fundamental research, e.g. for magnetization measurements of nanoscale particles and single spin detection \cite{schwarz2015low}. Moreover, GDBs could open the way to a range of future applications, such as HTS rapid single flux quantum (RSFQ) circuits \cite{wolf2013ybco} and  high-performance high-frequency HTS superconducting quantum interference filters (SQIFs) \cite{mitchell20162d}.} 
 
\begin{acknowledgement}

The authors acknowledge helpful discussions with T. Claeson and D. Montemurro. This work was been supported in part by the Knut and Alice Wallenberg Foundation (KAW) and in part by the Swedish Research Council (VR). R. A. is supported by the Swedish Research Council (VR) under the project ''Evolution of nanoscale charge order in superconducting YBCO nanostructures''. 

\end{acknowledgement}
 
\begin{suppinfo}

The following files are available.
\begin{itemize}
  \item Supporting Information: Ion milling parameters summary, current voltage characteristic for SQ2, differential resistance and modulation depth as a function of bias current for both SQ1 and SQ2.
\end{itemize}

\end{suppinfo}
 
\bibliography{bibliography}

\end{document}